\documentclass[letterpaper,titlepage,11pt]{article}
\usepackage{hyperref}
\usepackage{amssymb,amsmath,amsfonts,enumerate}
\usepackage{graphicx}

\def\clock{{\count0=\time
           \divide\count0 60
           \ifnum\count0<10 0\fi\the\count0
           \multiply\count0 -60 \advance\count0 \time
           :\ifnum\count0<10 0\fi \the\count0
         }}
\newcommand{\timestamp}{{\small\vbox{\hbox{\tt\jobname.tex}
\hbox{\the\day/\the\month/\the\year, \clock}}}}

\newcommand{\beq}{\begin{equation}}
\newcommand{\eeq}{\end{equation}}
\newcommand{\ben}{\begin{displaymath}}
\newcommand{\een}{\end{displaymath}}
\newcommand{\beqa}{\begin{eqnarray}}
\newcommand{\eeqa}{\end{eqnarray}}
\newcommand{\bea}{\begin{eqnarray}}
\newcommand{\eea}{\end{eqnarray}}
\newcommand{\bean}{\begin{eqnarray*}}
\newcommand{\eean}{\end{eqnarray*}}
\newcommand{\ba}{\begin{array}}
\newcommand{\ea}{\end{array}}
\newcommand{\bi}{\begin{itemize}}
\newcommand{\ei}{\end{itemize}}
\newcommand{\ie}{{\it i.e.,\,}}
\newcommand{\eg}{{\it e.g.,\,}}

\newcommand{\bbr}[1]{\mbox{${\mathbb R}^{#1}$}}

\numberwithin{equation}{section}

\begin{document}

\begin{titlepage}
\begin{flushright}
\end{flushright}
\vskip 2.cm
\begin{center}
{\bf\LARGE{Black Brane Viscosity and}}
\vskip 0.12cm
{\bf\LARGE{the Gregory-Laflamme Instability}} 
\vskip 1.5cm
{\bf Joan Camps$^{a}$, Roberto Emparan$^{a,b}$, Nidal Haddad$^{a}$
}
\vskip 0.5cm
\medskip
\textit{$^{a}$Departament de F{\'\i}sica Fonamental and}\\
\textit{Institut de
Ci\`encies del Cosmos, Universitat de
Barcelona, }\\
\textit{Mart\'{\i} i Franqu\`es 1, E-08028 Barcelona, Spain}\\
\smallskip
\textit{$^{b}$Instituci\'o Catalana de Recerca i Estudis
Avan\c cats (ICREA)}\\
\textit{Passeig Llu\'{\i}s Companys 23, E-08010 Barcelona, Spain}\\

\vskip .2 in
\texttt{jcamps@ub.edu, emparan@ub.edu, nidal@ffn.ub.es}

\end{center}

\vskip 0.3in

\baselineskip 16pt
\date{}

\begin{center} {\bf Abstract} \end{center} 

\vskip 0.2cm 

\noindent We study long wavelength perturbations of neutral black
$p$-branes in asymptotically flat space and show that, as anticipated in
the blackfold approach, solutions of the relativistic hydrodynamic
equations for an effective $p+1$-dimensional fluid yield solutions to
the vacuum Einstein equations in a derivative expansion. Going beyond
the perfect fluid approximation, we compute the effective shear and bulk
viscosities of the black brane. The values we obtain saturate generic
bounds. Sound waves in the effective fluid are unstable, and have been
previously related to the Gregory-Laflamme instability of black
$p$-branes. By including the damping effect of the viscosity in the
unstable sound waves, we obtain a remarkably good and simple
approximation to the dispersion relation of the Gregory-Laflamme modes,
whose accuracy increases with the number of transverse dimensions. We
propose an exact limiting form as the number of dimensions tends to
infinity.

\end{titlepage} \vfill\eject

\setcounter{equation}{0}

\pagestyle{empty}
\small
\normalsize
\pagestyle{plain}
\setcounter{page}{1}

\newpage

\section{Introduction and Summary}

Black holes exhibit thermodynamic behavior, so it is natural to expect
that their long wavelength fluctuations, relative to a suitable length
scale, can be described using an effective hydrodynamic theory. Over the
years there have appeared several different realizations of this idea,
which differ in the precise set of gravitational degrees of freedom that
are captured hydrodynamically (\eg only those inside a (stretched) horizon
as in \cite{membrane}, or the entire gravitational field up to a
large distance from a black brane spacetime as in
\cite{Bhattacharyya:2008jc,Emparan:2009at}) or in the kind of
asymptotics (Anti-deSitter \cite{Bhattacharyya:2008jc} or flat
\cite{Emparan:2009at}) of the black hole/brane geometry.

In this paper we focus on the hydrodynamic formulation developed recently for
higher-dimensional black holes, including asymptotically flat vacuum
black holes and black branes \cite{Emparan:2009at}. In this approach the
effective stress tensor of the `black brane fluid' is the quasilocal
stress tensor computed on a surface $B$ in a region that is
asymptotically flat in directions transverse to the brane\footnote{In
the following, asymptotic flatness always refers to directions
transverse to the brane.}. The equations of stress-energy conservation
describe both hydrodynamic (intrinsic) fluctuations along the
worldvolume of the brane, and elastic (extrinsic) fluctuations of the
brane worldvolume inside a `target' spacetime that extends beyond $B$.
Thus the dynamics of a black $p$-brane takes the form of the dynamics of
a fluid that lives on a dynamical worldvolume. This is referred to as
the blackfold approach.

In this paper we only study the intrinsic, hydrodynamic aspects of the
brane. The worldvolume geometry, defined by the surface $B$ at 
spatial infinity, is kept flat and fixed. Fluctuations of the
worldvolume geometry are non-normalizable modes, so the extrinsic
worldvolume dynamics decouples. With this simplification, the set up is
very similar to the fluid/AdS-gravity correspondence of
\cite{Bhattacharyya:2008jc}, which
we follow in many respects. The main difference is that we consider
vacuum black brane solutions, with no cosmological constant and with
different asymptotics. 

The quasilocal stress
tensor of a neutral vacuum black brane, with geometry equal to the
$n+3$-dimensional Schwarzschild-Tangherlini solution times $\bbr{p}$, is that
of a perfect
fluid with energy density $\rho$ and pressure $P$ related by the
equation of state
\beq\label{eqstate}
P=-\frac{\rho}{n+1}\,.
\eeq
We may choose the black brane temperature $T$ as the variable that
determines $\rho$ and $P$. The brane could also be boosted and thus have
a non-zero velocity field along its worldvolume. In a stationary
equilibrium state, the temperature and the velocity are uniform. We
study fluctuations away from this state where these quantities vary
slowly over the worldvolume. Their wavelength is measured relative to
the thermal length $T^{-1}$, so for a fluctuation with wavenumber $k$
the small expansion parameter is 
\beq
\frac{k}{T}\ll 1\,.
\eeq
Since for a vacuum black brane the temperature is inversely proportional
to the thickness of the brane, $r_0$, this can be equivalently expressed
as $k r_0\ll 1$.

To leading order in this expansion we obtain the hydrodynamics of an
effective perfect fluid, which
refs.~\cite{Emparan:2009cs,Emparan:2009at,Emparan:2009vd} have used to
derive non-trivial results for higher-dimensional black holes. At the
next order the stress tensor includes dissipative terms. For the purely
intrinsic dynamics, these are the shear and bulk viscosities, $\eta$ and
$\zeta$. In contrast to \cite{Bhattacharyya:2008jc}, our fluid is not
conformally invariant so $\zeta\neq 0$ is expected.

By analyzing long wavelength perturbations of the black brane and their
effect on the stress tensor measured near spatial infinity we
obtain
\beq\label{visco}
\eta=\frac{s}{4\pi}\,,\qquad \zeta=2\eta\left(\frac{1}{p}-c_s^2\right)
\eeq
where $s$
is the entropy density of the fluid, \ie $1/4G$ times the area
density of the black brane, and
\beq\label{soundspeed}
c_s^2=\frac{dP}{d\rho}=-\frac{1}{n+1}
\eeq
is the speed of sound, squared. 

Written in the form \eqref{visco}, these values for $\eta$ and $\zeta$
saturate the bounds proposed in \cite{Kovtun:2004de} and
\cite{Buchel:2007mf}. The result for the shear viscosity is not too
surprising: $\eta$ can be argued to depend only on the geometry near the
horizon and its ratio to $s$ is universal for theories of two-derivative
Einstein gravity \cite{Kovtun:2004de,Iqbal:2008by} (see also
\cite{Fujita:2007fg}). The bulk viscosity,
instead, does depend strongly on
the radial profile transverse to the brane\footnote{For instance, in the
membrane paradigm the bulk viscosity on the stretched horizon for a
generic black hole turns out to be negative. Our result \eqref{visco} is
instead positive.} so the saturation of the bound is presumably less
expected. Note, however, that these black branes have different
asymptotics than in all the previous instances where the effective
viscosities of black branes have been considered. In particular, these
black branes presumably are not dual to the plasma of any (local)
quantum field theory. In any case it is worth emphasizing that our
computations are for the theory with the simplest gravitational
dynamics: $R_{\mu\nu}=0$. 

The imaginary speed of sound \eqref{soundspeed} implies that sound waves
along the effective black brane fluid are unstable: under a
density perturbation the fluid evolves to become more and more
inhomogenous. Since this means that the black brane horizon itself
becomes inhomogeneous, ref.~\cite{Emparan:2009at} related this effect to the
Gregory-Laflamme (GL) instability of black branes
\cite{Gregory:1993vy}\footnote{This
connection had also been made for black branes with gauge
theory duals in \cite{Buchel:2005nt}.}. Then
\eqref{soundspeed} implies a simple form for the dispersion relation of
the GL unstable modes
$\omega(k)=-i\Omega(k)$
at long wavelength: 
$\Omega=k/\sqrt{n+1}\, +
O(k^2)$, \ie the slope of the curve $\Omega(k)$ near $k=0$ is exactly (and very
simply) determined
in the unstable-perfect-fluid approximation. 

\begin{figure}[ht]
\begin{tabular}{cc}
\includegraphics[width=0.5\textwidth]{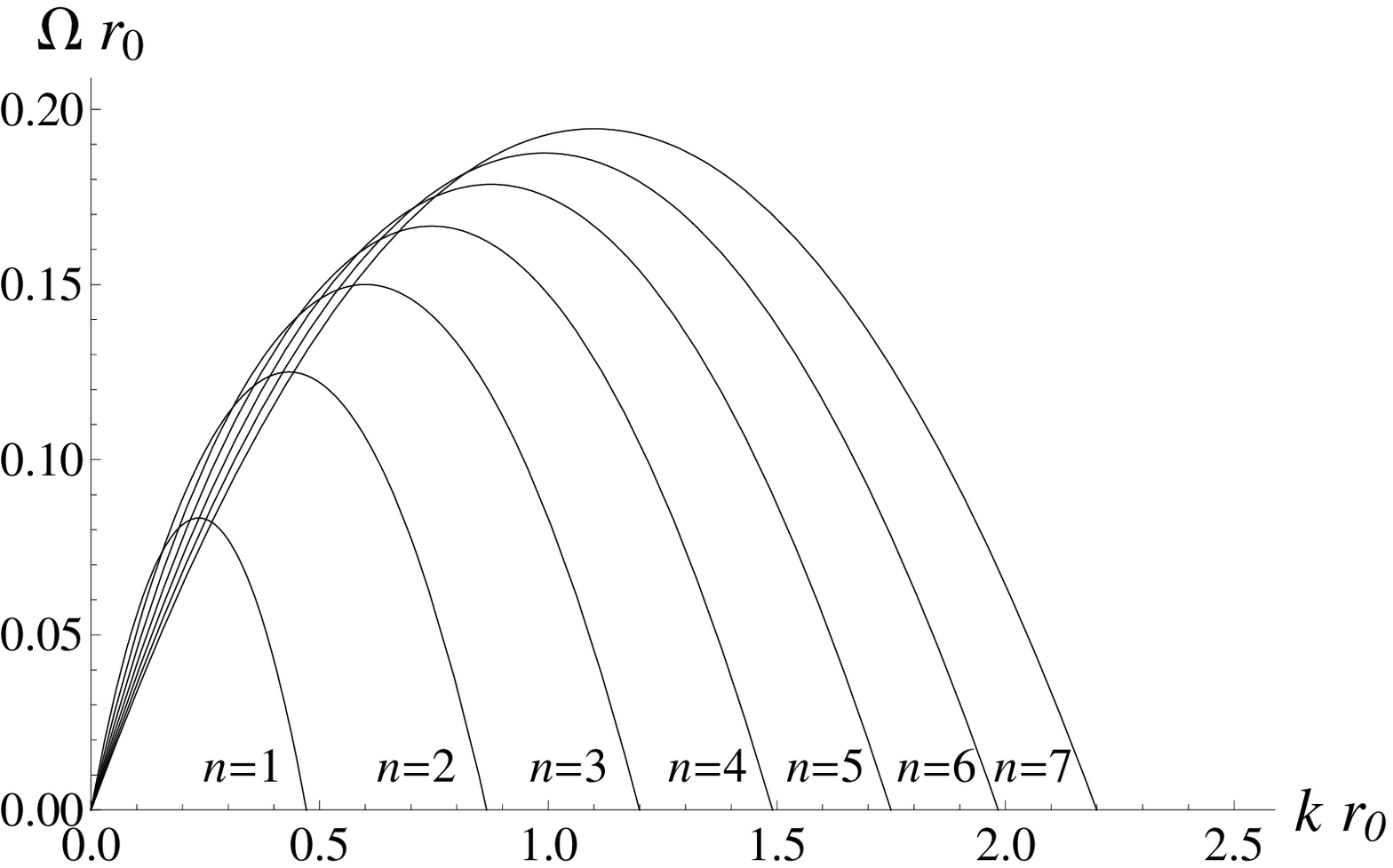}~
\includegraphics[width=0.5\textwidth]{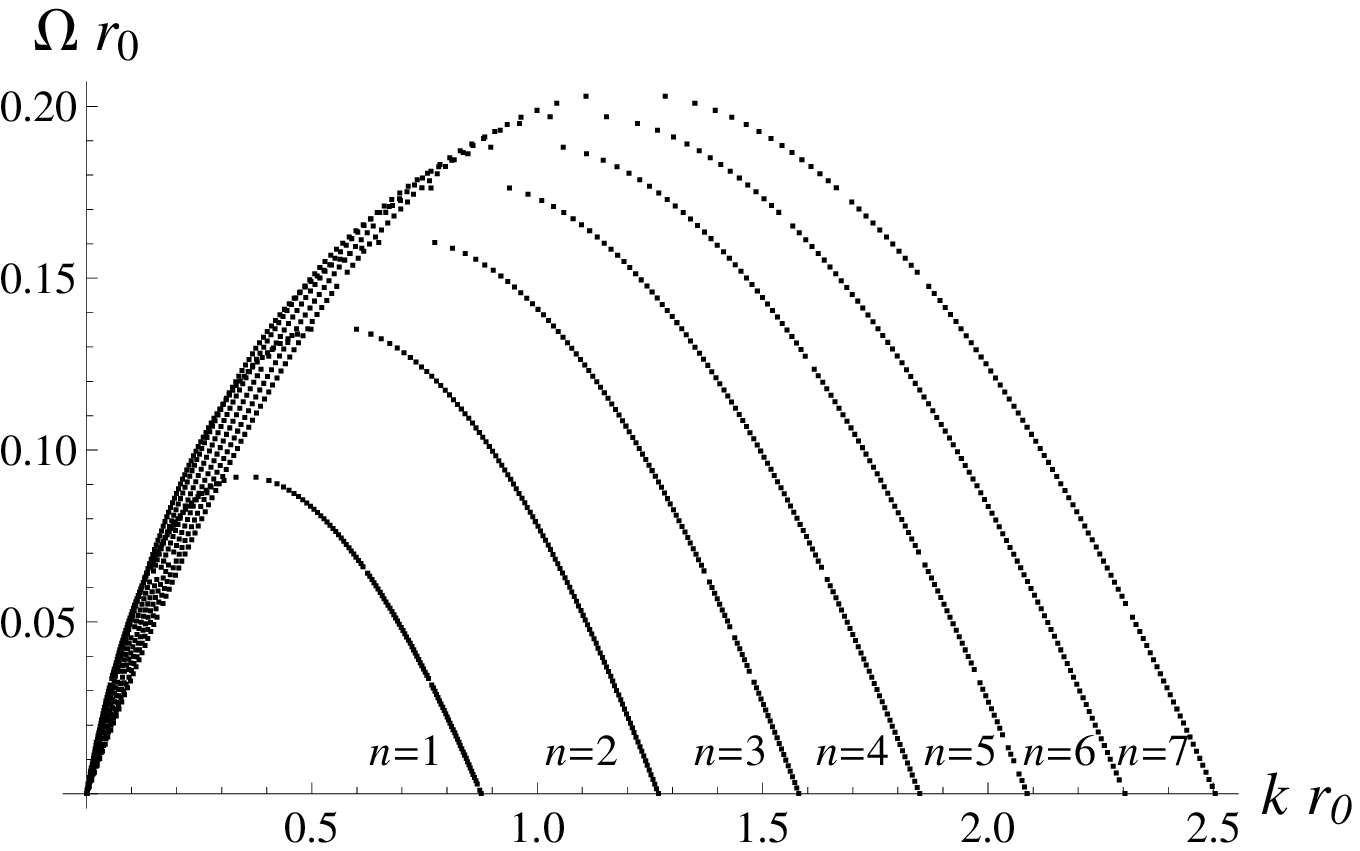}
\end{tabular}
\caption{Left: dispersion relation $\Omega(k)$, eq.~\eqref{GLvisc}, for
unstable sound waves in the effective black brane fluid (normalized
relative to the thickness $r_0$). Right:
$\Omega(k)$ for the unstable Gregory-Laflamme mode
for black branes (numerical data courtesy of P.~Figueras). For black
$p$-branes in $D$ spacetime dimensions, the curves depend only on
$n=D-p-3$. 
}\label{fig:compGL}
\end{figure}

Using our results for $\eta$ and $\zeta$ we can include
the viscous damping of sound
waves in the effective black brane fluid. The dispersion relation of
unstable modes becomes
\beq\label{GLvisc}
\Omega=\frac{k}{\sqrt{n+1}}\left(1-\frac{n+2}{n\sqrt{n+1}}\,k r_0\right)\,,
\eeq
which is valid up to corrections $\propto k^3$. 
Figure~\ref{fig:compGL} compares this
dispersion relation to the numerical results obtained from
linearized perturbations of a black $p$-brane. 
\begin{figure}[ht]
\begin{center}
\includegraphics[width=.95\textwidth]{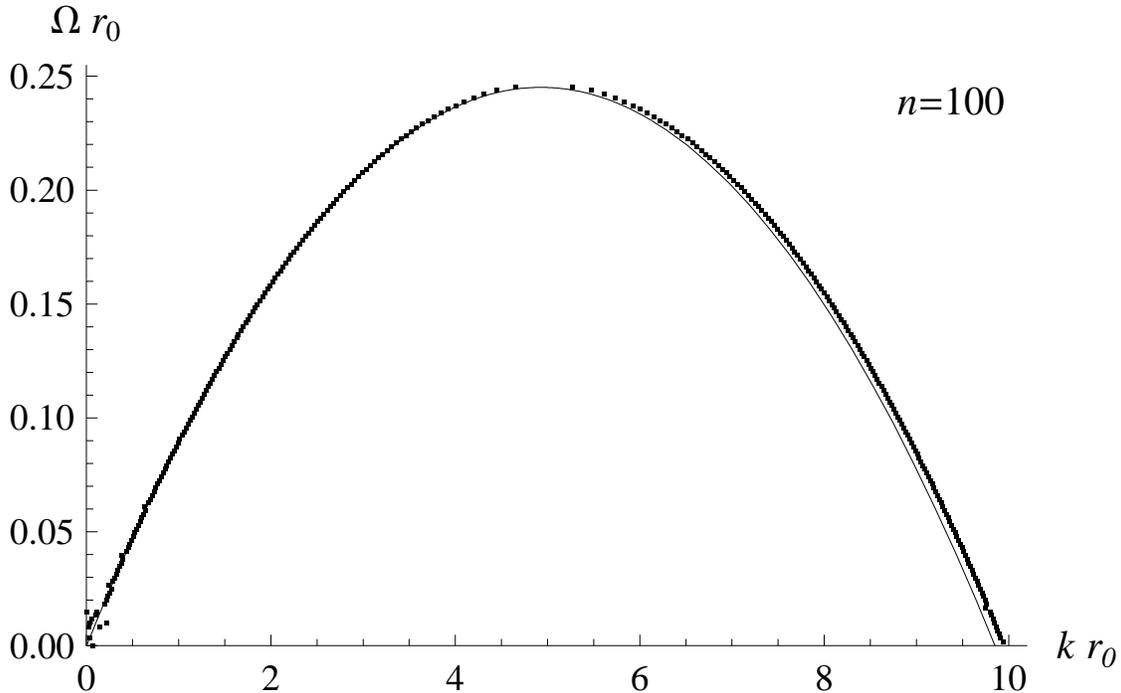} 
\end{center}
\caption{Dispersion
relation $\Omega(k)$ of unstable modes for $n=100$: the solid line is
our analytic approximation eq.~\eqref{GLvisc}; the dots are the
numerical solution of the Gregory-Laflamme perturbations of black branes
(numerical data courtesy of P.~Figueras).} 
\label{fig:compGL100}
\end{figure}
Zooming in on small values of $k r_0$, the match is excellent.
When $kr_0$
is of order one we have no right to expect agreement, but
the
overall qualitative resemblance of the curves is nevertheless striking. The
quantitative
agreement improves with increasing $n$ and indeed, as
figure~\ref{fig:compGL100} shows, at large $n$ it
becomes impressively
good over all wavelengths: for $n=100$ the
numerical values are reproduced to better than $1\%$ accuracy
up to the maximum value of $k$. Although the extent of this agreement is
surprising, we will provide some arguments for why
the fluid approximation appears to be so successful as $n$ grows.

Thus, the effective viscous fluid seems to capture in a simple manner
some of the most characteristic features of black brane dynamics. We
believe this is a significant simplification from the complexity of the
full Einstein equations.

\medskip

The outline of the rest of the paper is as follows: the next section
contains the bulk of the calculations of the paper for a generic
hydrodynamic-type perturbation of the black brane. We highlight the
differences with the analysis of \cite{Bhattacharyya:2008jc}, in
particular at asymptotic infinity, and compute the values \eqref{visco}
for the effective $\eta$ and $\zeta$. Section~\ref{sec:soundGL} relates
the linearized damped sound-mode perturbations of the fluid to the
Gregory-Laflamme perturbations of the black brane. We examine the
conditions that can lead to the surprising quantitative agreement of the
dispersion relation at large $n$, and we propose its exact form as
$n\to\infty$. We close in section~\ref{sec:discuss} with an examination
of the differences with other fluid-like approaches to the GL
instability, and a discussion of our results within the context of the
blackfold approach.

\section{Hydrodynamic perturbations of black branes}
\label{sec:hydropert}

In this section we study general perturbations of a vacuum black
$p$-brane with slow variation along the worlvolume directions of the
brane. Up to gauge transformations, they are fully determined by the
boundary conditions of horizon regularity and asymptotic flatness at
spatial infinity. Most of our analysis is very close to the study of
hydrodynamic perturbations of AdS black branes, but there is an additional
complication in the study of the perturbations at asymptotic infinity.
Nevertheless, we are able to find the complete explicit form of the
perturbed solution for a generic hydrodynamic flow to first
order in the derivative expansion. 

Readers who do not need or want the technical details of the calculation
of the perturbed solution and the viscous stress tensor can safely
skip to section~\ref{sec:soundGL}.

\subsection{Preliminaries}

\subsubsection{Black branes and their effective stress tensor}

The black $p$-brane solution of vacuum gravity in $D=p+n+3$ dimensions is
\beq\label{pbrane}
ds^2=\left(\eta_{ab}+\frac{r_0^n}{r^n}u_a u_b\right)d\sigma^a d\sigma^b
+\frac{dr^2}{1-\frac{r_0^n}{r^n}}+r^2 d\Omega_{n+1}^2\,,
\eeq
with $a=0,1,\dots,p$. The solution is characterized by the horizon radius $r_0$
(or brane `thickness') and the worldvolume velocity $u^a$, with $u^a
u^b\eta_{ab}=-1$. It is asymptotically flat in the directions transverse
to the
worldvolume coordinates $\sigma^a$. We can associate to it a
stress-energy tensor measured at spatial infinity. There are several
possible definitions of this stress tensor that would be equivalent for
calculational purposes, but for conceptual reasons the most convenient
for us is the quasilocal one of Brown and York \cite{Brown:1992br}. We
consider a boundary surface at large
constant $r$, with induced metric $h_{\mu\nu}$ and compute
\begin{equation}\label{intstress}
T_{\mu\nu}^\mathrm{(BY)}=\frac{1}{8\pi G}\left(K_{\mu\nu}-h_{\mu\nu}K -
(K^{(0)}_{\mu\nu}-h_{\mu\nu}K^{(0)})\right)\,,
\end{equation}
where $K_{\mu\nu}$ is the extrinsic curvature of the 
surface and we perform a
background substraction from flat
spacetime. 

The geometry of the boundary surface for \eqref{pbrane} is
$\bbr{1,p}\times S^{n+1}$.
We will introduce perturbations with wavelengths much longer
than the size $r_0$ of the $S^{n+1}$ at the horizon. The
deformations of this sphere all have large masses $\sim 1/r_0$ and
therefore decouple. Thus the
$SO(n+2)$ symmetry of $S^{n+1}$ is preserved and, in an appropriate
gauge, the metric will
remain a direct product with a factor of this sphere. We integrate over
the sphere to obtain the stress tensor for the black $p$-brane
\beq
T_{ab}=\int_{S^{n+1}}T_{ab}^\mathrm{(BY)}\,.
\eeq
	
We regard this stress tensor as living on the worldvolume of the brane,
\ie the $p+1$ extended directions of the boundary. The worldvolume metric
results from the asymptotic form of the boundary metric, which in our
case is the Minkowski metric
\beq
h_{ab}\to\eta_{ab}\,.
\eeq

A main advantage of using the
quasilocal stress tensor is that the Gauss-Codacci equations for the
constant-$r$ cylinder imply
$\partial^aT_{ab}\propto {R^r}_b$, so imposing the Einstein equations in
vacuum it follows that the stress tensor is conserved
\begin{equation}\label{stcons}
\partial^a T_{ab}=0\,.
\end{equation}

The stress tensor for the solution \eqref{pbrane} has the perfect fluid form
\begin{equation}\label{PerfFluid}
T_{ab}=\rho u_a u_b +P P_{ab}\,,\qquad
P_{ab}=\eta_{ab}+u_a u_b\,
\end{equation}
with energy density and pressure
\begin{equation}
\rho=-(n+1)P=(n+1)\frac{\Omega_{n+1}r_0^n}{16\pi G}\,.
\end{equation}
The horizon area allows to associate a local entropy density to this
effective fluid
\beq\label{entrdens}
s=\frac{\Omega_{n+1}r_0^{n+1}}{4 G}
\eeq
and all the thermodynamic functions can be expressed as functions of
the temperature
\beq
T=\frac{n}{4\pi r_0}\,.
\eeq
We can equivalently use $T$ or $r_0$ as the variable that
determines local equilibrium. In this section we will mostly use $r_0$
for notational simplicity.

We will be interested in preserving regularity at the horizon. This is
manifest if instead of the Schwarzschild
coordinates in \eqref{pbrane} we use Eddington-Finkelstein (EF) ones,
\begin{equation}\label{EFtoSchw0}
\sigma^a\rightarrow \sigma^a-u^a r_{*}\,,
\quad\quad r_{*}=\int\frac{1}{1-(r_0/r)^n}dr\,,
\end{equation}
such that
\begin{equation}
ds^2=-\left(1-\frac{r_0^n}{r^n}\right)u_a u_b
d\sigma^a d\sigma^b-2u_a
d\sigma^a dr+P_{ab}d\sigma^a d\sigma^b+r^2d\Omega_{n+1}^2\,.
\end{equation}

\subsubsection{Perturbations}

We promote the thickness and velocity parameters to collective fields
over the worldvolume, so
\beqa\label{EFpromoted}
ds_{(0)}^2&=&
-\left(1-\frac{r_0(\sigma)^n}{r^n}\right)u_a(\sigma)u_b(\sigma)
d\sigma^a d\sigma^b-2u_a(\sigma) d\sigma^a
dr+\left(\eta_{ab}+
u_{a}(\sigma)u_{b}(\sigma)\right)d\sigma^a d\sigma^b\nonumber\\
&&+r^2d\Omega_{n+1}^2
\,,
\eeqa 
where $r_0(\sigma)$ and $u^a(\sigma)$ are assumed to vary slowly relative
to the scale set by
$r_0$. In this paper we expand them to
first order in
derivatives, which we keep track of through a formal derivative-counting
parameter $\epsilon$. With non-uniform $r_0$ and $u^a$, the metric
\eqref{EFpromoted} is not Ricci flat so we add to it a
component with radial dependence
\begin{equation}\label{pertmetric}
ds^2=ds_{(0)}^2+\epsilon f_{\mu\nu}(r)dx^\mu dx^\nu+O(\epsilon^2)\,.
\end{equation}
We choose a gauge in which $\partial_r$ is a null
vector with normalization
fixed by the radius $r$ of $S^{n+1}$, so that
\begin{equation}
f_{rr}=0\,,\qquad f_{\Omega \mu}=0\,.
\end{equation}
With this choice the sphere $S^{n+1}$ can be integrated out.

Demanding that \eqref{pertmetric} satisfies the vacuum Einstein
equations to first order in $\epsilon$ results into a set of ODEs for
$f_{\mu\nu}(r)$. These will be
solved subject to regularity at the horizon $r=r_0$, which is easily
imposed as a condition of metric finiteness in EF
coordinates, and to asymptotic flatness, to which we turn next.

\subsubsection{Asymptotic infinity}

The asymptotic behavior of our spacetimes introduces an important
difference relative to the perturbations of AdS black branes.
For the latter, the calculations can be performed in their entirety in EF
coordinates in which
$\partial_r$ is a null vector. By taking large values of $r$ in these
coordinates one approaches
null infinity, but in AdS this is the same as spatial infinity. The
AdS boundary is always a timelike surface. However, in our asymptotically
flat space, null and spatial infinities differ. 

We are ultimately interested in computing the quasilocal stress tensor
on a timelike boundary of spacetime endowed with a non-degenerate
metric. But if we approach null infinity, the boundary metric will be
degenerate and it is unclear whether the quasilocal stress tensor is
well defined there. Instead, it seems more appropriate (and is
definitely unproblematic) to compute the stress tensor at spatial
infinity\footnote{Presumably the appropriate notion of spatial infinity here is not
Penrose's $i^0$ (which is just a point) but more along the lines of
\cite{Ashtekar:1991vb}, which naturally allows a dependence
along the boundary directions. Although our spatial infinity is not
exactly the same as in \cite{Ashtekar:1991vb} since instead of a
hyperboloid we work on a cylinder where $\bbr{1,p}$ and $S^{n+1}$ scale
differently at infinity, this is not a problem for us since we are
integrating over $S^{n+1}$. It would be interesting, especially with a
view to holography, to further formalize this notion of spatial
infinity. Related remarks concerning holography in
asymptotically flat spacetimes have been made in \cite{Marolf:2006bk}.}.
For this purpose EF coordinates are very awkward and it is much more
convenient to switch back to Schwarzschild-like coordinates
$\{r,t,\sigma^i\}$ at large $r$. 

Thus we will work with two sets of coordinates: EF ones, in which
horizon regularity is manifest, and Schwarzschild coordinates, in which
spatial infinity is naturally approached. We need to provide the change
of coordinates that relates them, extending the inverse of
\eqref{EFtoSchw0} to include $O(\epsilon)$ terms. The correction is
naturally guessed by recalling that $u^a$ and $r_0$, which appear in the
transformation \eqref{EFtoSchw0}, now depend on the EF coordinates.
Thus,
\begin{equation}\label{EFtoSchw}
\sigma^a\rightarrow \sigma^a+u^a(v,\sigma^i) 
\int\frac{dr}{1-(r_0(v,\sigma^i)/r)^n}\,,
\end{equation}
or more explicitly,
\beqa\label{EFtoSchw2}
v&\rightarrow& t+r_{*}+\epsilon\frac{(t+r_*)\partial_v
r_0+\sigma^i\partial_i
r_0}{r_0}\left(r_*-\frac{r}{1-(r_0/r)^n}\right)+O(\epsilon^2)\,,\nonumber\\
\sigma^{i}&\rightarrow&\sigma^{i}+\epsilon 
\left((t+r_*)\partial_v u^{i}+\sigma^j\partial_j u^i\right)r_{*}+O(\epsilon^2)\,.
\eeqa

\subsection{Solving the perturbation equations}

At each point we choose coordinates centered on that point and go to an
(unperturbed)
local rest frame. In
EF coordinates the velocity
perturbation is
\beq
u^v(\sigma)=1+O(\epsilon^2)\,,\qquad
u^i(\sigma)=\epsilon\sigma^{a}\partial_{a}u^i(0)+O(\epsilon^{2})\,.
\eeq
Note that since local velocities are small the constraint $u^2=-1$ is
automatically satisfied to the order we need.
The other collective variable of the effective black brane fluid is the
temperature $T$, or equivalently the thickness $r_0$, which we perturb as
\beq
r_0(\sigma)=r_0(0)+\epsilon
\sigma^{a}\partial_{a}r_0(0)+O(\epsilon^{2})\,.
\eeq
In the following we understand all quantities as evaluated at
$\sigma^a=0$ and thus denote $\partial_{a}u^i(0)\to \partial_{a}u^i$,
$r_0(0)\to r_0$ etc.

The metric \eqref{pertmetric} is now
\begin{align}
ds^2=&2dv dr-f(r)dv^2+
\sum_{i=1}^{p}d\sigma_{i}^{2}+r^2d\Omega_{n+1}^{2}\notag\,\\ 
&-2\epsilon\sigma^{a}\partial_{a}u_{i}d\sigma^{i}dr+
\epsilon\frac{nr_{0}^{n-1}\sigma^{a}\partial_{a}r_0}{r^{n}}dv^2-
2\epsilon\frac{r_0^{n}\sigma^{a}\partial_{a}u_{i}}{r^{n}}d\sigma^{i}d
v
+\epsilon f_{\mu\nu}(r)dx^{\mu}dx^{\nu}\,,
\end{align}
where we denote
\beq
f(r)=1-\frac{r_0^n}{r^n}\,.
\eeq

The Einstein equations with a radial index, ${R^r}_a=0$ do not involve
second derivatives and are constraint equations. Indeed they only
involve the hydrodynamic fields $r_0$ and $u^i$ and not $f_{\mu\nu}$,
\beq
(n+1)\partial_{v}r_0=-r_0\partial^{i}u_{i}\,,\qquad
\partial_{i}r_0=r_0\partial_{v}u_{i}\label{eoms},
\eeq
so they are to be regarded as the equations of fluid dynamics,
consistently with \eqref{stcons}. We also
verify this interpretation later. 

The remaining
Einstein's equations are dynamical and we solve them to find
$f_{\mu\nu}$.
The equations $R_{ij}=0$ give
\begin{equation}
\partial_{r}\left(r^{n+1}f
{f_{ij}}'\right)=-2(n+1)r^{n}\partial_{(i}u_{j)}\,,
\end{equation} 
which, requiring finiteness at the horizon, are solved by
\begin{equation}
f_{ij}(r)=c_{ij}-2\partial_{(i}u_{j)}\left(r_{*}-
\frac{r_0}{n}\log f\right)\,.
\end{equation} 
The integration constants $c_{ij}$ will be fixed later demanding asymptotic
flatness. The equations $R_{v i}=0$,
\begin{equation}
\partial_{r}\left(r^{n+1}{f_{v
i}}'\right)=-(n+1)r^{n}\partial_{v}u_{i}\,,
\end{equation}
are solved by
\begin{equation}
f_{v i}=c_{v i}^{(2)}+\frac{c_{v
i}^{(1)}}{r^{n}}-\partial_{v}u_{i}r\,,
\end{equation}
which are regular at the horizon for all
values of the constants. 
Next, the equations from 
$R_{rr}=0$ and $R_{\Omega\Omega}=0$ are
\begin{equation}
{f_{v r}}'=\frac{r}{2(n+1)}\sum_{i=1}^{p}{f_{ii}}''\,,
\end{equation}
and
\begin{equation}
\partial_{r}\left(r^{n}f_{v
v}\right)=
r^{n}\partial^{i}u_{i}+\frac{r^{n}f}{2}\left(\sum_{i=1}^{p}{f_{ii}}'-
2{f_{v r}}'\right)-2nr^{n-1}f_{v r}\,,
\end{equation}
which, assuming that eqs.~\eqref{eoms} are satisfied, are solved by
\begin{equation}
f_{vr}=c_{vr}+\frac{r^2}{2(n+1)}\frac{d}{dr}\sum_i\frac{f_{ii}}{r}\,,
\end{equation}
and
\begin{equation}
f_{v
v}=\frac{2\partial^{i}u_{i}r+\left(1-
\frac{n+2}{2}\frac{r_0^n}{r^{n}}\right)\sum_{i=1}^{p}f_{ii}}{n+1}-2c_{v
r}+\frac{r_0^n}{r^{n}}c_{vv}\,.
\end{equation}
Again these are regular at the horizon for all choices of the
integration constants.
Note that $f_{rj}$ does not appear in Einstein's equations to first
order in $\epsilon$ and corresponds to a gauge
mode. This, and the integration constants, will be fixed shortly. 

At this stage, for any hydrodynamic perturbation that solves
the equations \eqref{eoms}, we have managed to construct a perturbed
metric that is regular at the horizon. Next we must ensure that the
solution remains asymptotically flat.
Transforming to 
Schwarzchild-like coordinates using \eqref{EFtoSchw2}, we require that 
\beq
g_{ab}=\eta_{ab}+O(r^{-n})\,.
\eeq
For the other metric components, we find that
$g_{rr}=1+O(r^{-n})$, $g_{ri}=O(r^{-n})$, and
$g_{tr}=O(r^{-n+1})$ when $n>1$ ($g_{tr}=O(\log r/r)$ when $n=1$),
are enough to obtain a finite stress tensor. Recall
also that all the metric components involving angular coordinates of
$S^{n+1}$ are unaltered.

Omitting details, we find that the conditions on
$g_{ij}$ and $g_{tj}$ fix
\beq
c_{ij}=c_{v j}^{(2)}=0\,.
\eeq 
In addition, the effect of $c_{v j}^{(1)}$ in $g_{tj}$
amounts to a global shift in the velocity field
along the spatial directions of the brane, so in order to remain in a local
rest frame we set 
\beq
c_{v j}^{(1)}=0\,.
\eeq
Furthermore, if we perform the change
\begin{equation}
t\rightarrow t\left(1-\epsilon c_{v r} \right)\,,
\end{equation} 
then $c_{vv}-2c_{v r}$ results in a global shift in the
temperature, which we eliminate by choosing
\beq
c_{v
v}=2c_{v r}\,.
\eeq
Asymptotic flatness in $g_{tr}$ imposes a choice for $c_{v r}$
that singles out the slower fall-off of $n=1$,
\begin{equation}
c_{v r}=-\partial_{t}r_0\quad \textrm{for}~n=1\,,\qquad
c_{v r}=0\quad \textrm{for}~n>1\,
\end{equation}
(note that the values of $\partial_t r_0$ and $\partial_v r_0$ at
$\sigma^a=0$ are equal).

Asymptotic flatness in these coordinates is a little delicate when $n=1$
due to its slower fall-off, and to make it manifest we take an $f_{rj}$
gauge diverging at infinity. This is not necessary when $n>1$ (and
neither choice affects the calculation of the stress tensor).
Thus we set
\begin{equation}
f_{rj}=-\partial_{j}r_0 \log \frac{r}{r_0}\quad
\textrm{for}~n=1\,,\qquad
f_{rj}=0\quad \textrm{for}~n>1\,.
\end{equation}

Summarizing, we obtain
\begin{align}
g_{ij}=&\delta_{ij}+\epsilon r_0\frac{2\partial_{(i}u_{j)}}{n}\log f\,,\\
g_{t j}=&-\epsilon\frac{r_0^{n}}{r^n}\sigma^{a}\partial_{a}u_{j}\,,\\
g_{t r}=&\epsilon \frac{\partial_{t}r_0}{f}\left(\left(
\frac{r_0^n}{r^n}-\frac{f}{n}\right)\log f-\frac{ r_0^{n}}{r^n}
\left(n\frac{r_{*}}{r_0}+1\right)\right)-\epsilon c_{v
r}\,,\\
g_{r j}=&\epsilon f_{rj}(r)+\epsilon\frac{\partial_j
r_0}{r_0}\frac{r_*-r}{f}\,,\\
g_{r r}=&f^{-1}+\epsilon
f^{-2}\left(
\frac{nr_0^{n-1}\sigma^{a}\partial_{a}r_0}{r^n}+
\frac{r_0^n\partial_{t}r_0}{r^n}\left(\log f-2\right)\right)\,,\\
g_{t t}=&-f+\epsilon
\left(\frac{nr_0^{n-
1}\sigma^{a}\partial_{a}r_0}{r^n}+\partial_{t}r_0\log
f\left(\frac{r_0^n}{r^n}-\frac{2}{n}f\right)\right)\,,
\end{align}
($\sigma^a$ correspond to Schwarzschild coordinates here, so $\sigma^0=t$). This
is the complete solution for the black brane metric that corresponds to
a hydrodynamic perturbation that solves the equations \eqref{eoms}
expanded around the origin of the local rest frame, $\sigma^a=0$.

\subsection{Viscous stress tensor}

We are now ready to compute the quasilocal stress tensor
\eqref{intstress}. The renormalization via background
subtraction is simple and appropriate, since our metrics are
infinitesimally close to the uniform black
$p$-brane and their asymptotic boundaries can
always be embedded in flat spacetime. Straightforward calculations give
\beqa\label{corrstress}
T_{ij}&=&\frac{\Omega_{n+1}}{16\pi G}\left(-\delta_{ij}(r_{0}+
\epsilon\sigma^a\partial_a r_0)^{n}-
\epsilon r_0^{n+1}\left[\left(2\partial_{(i}u_{j)}-
\frac{2}{p}\delta_{ij}\partial^{\ell}u_{\ell}\right)+2\left(\frac{1}
{p}+\frac{1}{n+1}\right)\delta_{ij}\partial^{\ell}u_{\ell}\right]\right)\,,\nonumber\\
T_{t t}&=&\frac{\Omega_{n+1}}{16\pi G}(n+1)\left(r_0+\epsilon 
\sigma^{a}\partial_{a}r_0\right)^n\,,\\
T_{t j}&=&-\frac{\Omega_{n+1}r_{0}^{n}}{16\pi G} \epsilon
n\sigma^{a}\partial_{a}u_{j}\,,\nonumber
\eeqa
 which are valid up to $O(\epsilon^2)$. One can easily check that
the hydrodynamic equations $\partial_{a}T^{ab}=0$ are
indeed equivalent to the constraint equations \eqref{eoms}.

Write now this stress tensor in the form
\begin{equation}\label{Tdiss}
T_{ab}=\rho u_a u_b +P P_{ab}-\zeta \theta P_{ab}-2\eta
\sigma_{ab} + O(\partial^2)
\eeq
where the expansion and shear of the velocity congruence are
\beq
\theta=\partial_a u^a\,,\qquad 
\sigma_{ab}={P_a}^c\left(\partial_{\left(c\right.}u_{\left.d\right)}-
\frac{1}{p}P_{cd}\right){P^d}_b\,.
\end{equation}

The component $T_{tt}$ in \eqref{corrstress} determines the energy
density, and requiring that the
equation of state \eqref{eqstate} holds locally uniquely identifies the
pressure. Then we can write
\begin{equation}
T_{ij}=P\delta_{ij}-\epsilon\eta\left(2\partial_{(i}u_{j)}-
\frac{2}{p}\delta_{ij}\partial^{\ell}u_{\ell}\right)-
\epsilon\zeta\delta_{ij}\partial^{\ell}u_{\ell}
\end{equation}
with
\beq
\eta=\frac{\Omega_{n+1}}{16 \pi G }r_{0}^{n+1}\,,\qquad
\zeta=\frac{\Omega_{n+1}}{8 \pi G
}r_{0}^{n+1}\left(\frac{1}{p}+\frac{1}{n+1}\right)\,.
\eeq
Using \eqref{soundspeed} and \eqref{entrdens} these can be rewritten as
in \eqref{visco}.

\section{Damped unstable sound waves and the Gregory-Laflamme instability}
\label{sec:soundGL}

Our analysis in the previous section applies to generic long-wavelength
perturbations of arbitrarily large amplitude.
Let us now consider small perturbations of a static fluid of the form
\beq
\rho\to\rho+\delta\rho\,,\qquad P\to P+c_s^2 \delta \rho\,,\qquad
u^a=(1,0,\dots)\to (1,\delta u^i)\,,
\eeq
where
$c_s$ is the
speed of sound, and with 
\beq
\delta\rho(t,\sigma^i)=\delta\rho\, e^{i\omega t+ik_j\sigma^j}\,,
\qquad \delta u^i(t,\sigma^i)=\delta u^i\, e^{i\omega t+ik_j\sigma^j}\,.
\eeq
We substitute these in the viscous fluid equations and linearize in the
amplitudes $\delta\rho$ and $\delta u^i$, to find
\begin{eqnarray}
\omega\delta\rho+(\rho+P)k_i \delta u^i+ O(k^3)&=&0\,,\\
i\omega(\rho+P)\delta u^j+ic_s^2k^j\delta\rho+
\eta k^2 \delta u^j+k^j \Biggl(\left(1-\frac{2}{p}\right)\eta+
\zeta\Biggr)k_l \delta u^l+O(k^3)&=&0\,.
\end{eqnarray}
Applying our results above, any solution to these
equations
can be used to obtain an explicit black brane solution with a small,
long-wavelength
fluctuation of $r_0$ and $u^a$.
If we eliminate $\delta\rho$ we find that non-trivial sound waves require
\begin{equation}\label{omk}
\omega-c_s^2 \frac{k^2}{\omega}-i\frac{
k^2}{Ts}\Biggl(2\left(1-
\frac{1}{p}\right)\eta+\zeta\Biggr)+O(k^3)=0\,,
\end{equation}
where $k=\sqrt{k_ik_i}$ and we have used the Gibbs-Duhem relation $\rho+P=T s$. 
This equation determines the dispersion relation $\omega(k)$. For a
stable fluid with $c_s^2>0$, viscosity adds a small imaginary part
to the frequency, which becomes complex and describes damped sound
oscillations. Instead our effective fluid has imaginary sound-speed,
eq.~\eqref{soundspeed}, so $\omega$ is purely imaginary: sound waves are
unstable.
Writing
\beq
\omega=-i\Omega
\eeq
we solve \eqref{omk} to find
\begin{equation}
\Omega=\sqrt{-c_s^2}k-\Biggl(\left(1-
\frac{1}{p}\right)\frac{\eta}{s}+\frac{\zeta}{2s}\Biggr)
\frac{k^2}{T}+O(k^3)\,.
\end{equation}
For the specific black $p$-brane fluid this yields the dispersion
relation \eqref{GLvisc}. The connection between these unstable sound
waves and the Gregory-Laflamme instability was pointed out at the
perfect fluid level (\ie $\Omega$ linear in $k$) in
\cite{Emparan:2009at}, and we have discussed it in the
introduction.\footnote{Observe that the result \eqref{GLvisc} is independent of
$p$. That this must be the case is clear from the outset in the GL
analysis and also in our analysis of the Einstein equations.} 

Figures~\ref{fig:compGL} and \ref{fig:compGL100} show that our
approximation \eqref{GLvisc}
improves as $n$
grows. In order to see how this might be justified, let us first
rewrite the dispersion relation \eqref{GLvisc} in terms of the
temperature $T$ instead of $r_0$, 
\beq\label{GLviscT}
\Omega=\frac{k}{\sqrt{n+1}}\left(1-\frac{n+2}{\sqrt{n+1}}\,\frac{k}{4\pi T}
+O(k^2/T^2)\right)\,.
\eeq
In principle, at any given $n$, both quantities $r_0$ and $T^{-1}$
define length scales that are parametrically equivalent. But if we vary
$n$ and allow it to take large values, then $r_0$ and $T^{-1}\sim r_0/n$ can
differ greatly. We propose that in this case, $T^{-1}$, and not
$r_0$, is the length scale that limits the validity of the fluid
approximation, so the appropriate expansion variable for large $n$ is
$k/T$ and not $k r_0$. This may actually be natural since from
the fluid point of view $T$ has a clearer physical meaning than $r_0$. In
effect, we are proposing that when $n\gg 1$ it is more accurate to view
the effective
theory as describing very \textit{hot} black branes, rather than very
\textit{thin} ones.

The point of this exercise is that for large $n$ the maximum values over
which $\Omega$ and $k$ in \eqref{GLviscT} range are
$(k/T)|_\mathrm{max}\sim 1/\sqrt{n}$ and $(\Omega/T)|_\mathrm{max}\sim
1/n$. So as $n$ grows the frequency and wavenumber of unstable modes
extend over a smaller range of $k/T$ and $\Omega/T$. This strongly
suggests that hydrodynamics can capture more accurately the dynamics of
GL modes when the number of dimensions becomes very large.\footnote{This
is similar in spirit, although not precisely equal, to the proposal in
\cite{Caldarelli:2008mv} that in the limit of large number of dimensions
black holes are accurately described by fluid mechanics.} More
precisely, if we write the corrections inside the brackets in
\eqref{GLviscT} in the form $\sum_{j\geq 2} a_j (k/T)^j$, and assume
that the $n$-dependence of the coefficients $a_j$ is such that $a_j
n^{-j/2}\to 0$ as $n\to\infty$, then the expansion in $k/T$, \ie the
hydrodynamic derivative expansion, becomes a better approximation over a
larger portion of the curves $\Omega(k)$. 

This is a relatively mild-looking assumption
on the $n$-dependence of the higher-order coefficients in the expansion in
$k/T$,\footnote{Which, crucially, is not satisfied by the coefficient
of the linear term inside the brackets in \eqref{GLviscT}.} and in particular is
satisfied if the $a_{j\geq 2}$ remain finite as $n\to\infty$. But since
we
have not computed higher-derivative transport
coefficients then, within our
perturbative framework, we cannot prove its validity. 
However, since the numerical data appear to strongly support it, we
conjecture that the truncation of the dispersion relation
up to $k^2$-terms captures the
complete dispersion relation at
large $n$. More precisely, if we define a rescaled
frequency and wavenumber,
\beq\label{nscaling}
\tilde\Omega=n\Omega\,,\qquad \tilde k=\sqrt{n}k
\eeq
that remain finite as $n\to\infty$, then we propose that 
\beq\label{limOm}
\tilde\Omega=\tilde k\left(1-\frac{\tilde k}{4\pi T}\right)
\eeq
is the exact limiting relation valid for all wavenumbers $0\leq \tilde k\leq
4\pi T$.

Note that the truncation of $\Omega(k)$ in \eqref{GLviscT} appears to
capture the zero-mode with
$\Omega=0$ at a finite $k=k_{GL}$. This is quite remarkable, since the
viscous fluid equation
\eqref{omk} does not admit any zero-mode solution. 
The comparison with numerical data in figure~\ref{fig:compGL100} shows
that the quantitative result
for $k_{GL}$,
although poor for small $n$, becomes excellent for large $n$. 
Further evidence for the validity of our proposal
comes from the analytical
value of the GL zero
mode in the limit $n\to\infty$ \cite{Kol:2004pn}
\beq
k_{GL}\to \frac{4\pi T}{\sqrt{n}}\,.
\eeq
This is the same as the limiting value for the zero-mode
`predicted' by \eqref{nscaling}, \eqref{limOm}.\footnote{The
relative difference between the results for $k_{GL}$ from the large-$n$
subleading correction
computed in \cite{Asnin:2007rw} and from \eqref{GLvisc} is equal to
$1/n$. This is precisely the size of the
discrepancy observed in
fig.~\ref{fig:compGL100}.}

Presumably, by effecting the scaling \eqref{nscaling} in the
full linearized perturbation equations of the GL problem one may prove (or
possibly disprove) equation \eqref{limOm}.

\section{Discussion}
\label{sec:discuss}

Our analysis of the GL instability must not be confused with recent
studies where a connection to the Rayleigh-Plateau instability of fluid
tubes is made. In the latter approach, following a suggestion in
\cite{Cardoso:2006ks}, refs.~\cite{Caldarelli:2008mv,Maeda:2008kj}
related a $d$-dimensional black string in a Scherk-Schwarz
compactification of Anti-deSitter space to a $d-2$-dimensional fluid
tube with a boundary with surface tension (see
\cite{Aharony:2005bm}). The Rayleigh-Plateau instability of the fluid
tube arises from the competition between surface tension and bulk
pressure. In contrast, our effective fluid does not have any boundaries
so the instability is not of the Rayleigh-Plateau type, but rather one
in the sound modes. Also note that our calculations in
sec.~\ref{sec:hydropert} yield explicit black brane solutions to the
Einstein equations (in vacuum) in a derivative expansion, something
that, although expected to be possible in principle, at present cannot
be realized for the fluid solutions in
\cite{Caldarelli:2008mv,Maeda:2008kj}. 

We stress that our analysis is not a `dual' solution of
the GL instability problem: we have investigated the same perturbation problem
as in \cite{Gregory:1993vy} and explicitly solved it in closed analytic
form in a derivative expansion. Since our approach does not require the
perturbations to be small, it may even be used to study the non-linear
evolution of the GL instability.

One of our motivations has been to show explicitly how the effective
theory of blackfolds of \cite{Emparan:2009at} can be systematically
developed as a derivative expansion of the Einstein equations. Although
we have done it only for the intrinsic aspects of blackfold dynamics, we
have been able to: (i) derive in detail, starting from the `microscopic'
(full Einstein) theory, the lowest-order blackfold formalism that
ref.~\cite{Emparan:2009at} had developed following general principles;
(ii)~prove that the first corrections to the lowest-order formalism can
be computed and result in perturbations of the black brane that preserve
regularity of the horizon. The viscosity coefficients are determined
precisely from this condition.

In general, the worldvolume of a blackfold is dynamical and can be
curved. Our calculations in this paper can be regarded as being valid
for fluid perturbations with a wavelength that, while longer than
$T^{-1}$, is much shorter than the typical curvature radius $R$ of the
blackfold worldvolume. In this case, the intrinsic and extrinsic
dynamics decouple. Thus, for a curved blackfold our results for the GL
instability are valid at most up to wavelengths smaller than $R$. At
longer wavelengths the hydrodynamics of the effective
fluid is fully coupled to the elastic dynamics of the worldvolume. For
instance this is case for perturbations of thin black rings with
wavelength comparable to the ring radius. These lie beyond the range of
applicability of our results.

It should be quite interesting to extend our analysis to include the
extrinsic aspects of the blackfold. To do this, one first allows the
worldvolume metric where the fluid lives to be a curved background, with
an extrinsic curvature radius much larger than $T^{-1}$. This curvature
acts as an external force on the fluid \cite{Emparan:2009at}. In the
derivative expansion, the stress tensor will in general contain, besides
the viscosities, higher-derivative coefficients that multiply
derivatives of the worldvolume metric. These coefficients will be
determined by demanding horizon regularity of a perturbation that curves
the asymptotic geometry. Perturbations of this kind have been studied
for certain illustrative examples in
\cite{Emparan:2007wm,Caldarelli:2008pz,Emparan:2009vd} in stationary
situations that do not involve viscous dissipation. Thus it may be
possible to extract the extrinsic pressure coefficients in the stress
tensor.

In the AdS context, the external force on the fluid from a worldvolume
curvature has been studied in \cite{Bhattacharyya:2008ji}. However, in
that case the worldvolume geometry is regarded as a fixed, non-dynamical
background. Instead, in the blackfold context this geometry is
dynamical. A solution of the forced fluid equations will backreact on
the background spacetime where the blackfold lives, and thus modify the
worldvolume geometry. Therefore for a generic, curved blackfold the
explicit construction of perturbative metrics becomes rather more
complicated than in the fluid/AdS-gravity correspondence.

\section*{Acknowledgments}

We are indebted to Pau Figueras for kindly providing the excellent
numerical data for the Gregory-Laflamme instability used in
figures~\ref{fig:compGL} and \ref{fig:compGL100}. We thank Troels
Harmark, Vasilis Niarchos, and Niels Obers for discussions. JC thanks
the NBI for warm hospitality. Work supported by DURSI 2009 SGR 168, MEC
FPA 2007-66665-C02 and CPAN CSD2007-00042 Consolider-Ingenio 2010. JC
was also supported in part by FPU grant AP2005-3120.


\end{document}